\documentclass[9pt, twocolumn]{article}
\usepackage[T1]{fontenc}
\usepackage{amssymb,amsmath,bm}
\usepackage{theorem}
\usepackage{graphicx}
\usepackage[justification=centering]{caption}
\usepackage[utf8]{inputenc}
\def\BibTeX{{\rm B\kern-.05em{\sc i\kern-.025em b}\kern-.08em
    T\kern-.1667em\lower.7ex\hbox{E}\kern-.125emX}}

\newtheorem{definition}{Definition}

\newtheorem{proposition}{Proposition}
\newtheorem{example}{Example}
\newenvironment{propositionbis}
  {\addtocounter{proposition}{-1}%
   \begin{proposition}}
  {\end{proposition}}

\begin{document}
\title{Computing some Principal Value integrals without Residues and Applications on Hilbert Transform and Fourier Transform}
\author{Jorge Pedraza Arpasi
\thanks{Jorge Pedraza Arpasi
Centro Tecnologico de Alegrete, Universidade Federal do Pampa - UNIPAMPA Alegrete-RS Brasil,
E-mail: jorgearpasi@unipampa.edu.br.}}
\maketitle

\begin{abstract}
This article proposes a new approach in the treatment of the Hilbert transform and some cases of the Fourier transform
whose improper integrals are principal values.
This approach may be useful for teaching these issues to undergraduate engineering students.
Traditional literature of Complex Analysis deals with these transformation integrals with the Cauchy-Goursat theorem and the
residues calculation technique. In this new approach, instead of residues, we use an intuitive result about complex line integrals of continuous complex functions that resembles the delta of Dirac.

\vspace{1cm}

\noindent\textbf{\Large Resumo}

\noindent Neste artigo é proposto uma nova abordagem na manipulação da transformada de Hilbert e
alguns casos da transformada de Fourier
cujas integrais impróprias sejam  valores principais.
Esta abordagem pode ser de utilidade no ensino destes assuntos para estudantes de graduação em engenharias.
A literatura tradicional de Variável Complexa trata estas integrais de transformada com o teorema de Cauchy-Goursat e o cálculo
de resíduos. Em esta abordagem nova, ao invés de resíduos, usaremos um resultado intuitivo acerca de integrais de linha de funções
continuas complexas e que tem um certa semelhança com o delta de Dirac.

\noindent \textbf{Palavras Chave.-} Transformada de Hilbert, Transformada de Fourier, Teorema de Cauchy-Goursat.

\end{abstract}

\section{Introduction}
The Hilbert transform of a real function $f(x)$ is defined as the principal value of the integral
\begin{equation}\label{eq:hilbert}
 \mathcal{H}\{f\}(w)=\int\limits_{-\infty}^\infty \frac{f(x)}{w-x} dx
\end{equation}
In Signal Processing theory, the Hilbert transform is a fundamental tool to  provide the mathematical basis for transforming
real-valued band-pass signals and systems into their low-pass equivalent representations without loss of information
\cite{haykin,lathi}. The equation (\ref{eq:hilbert}) can be generalized in several ways, as the proposed by
\cite{zayed, brackx, caleb}. However, the Hilbert transform (\ref{eq:hilbert}) seems artificial, to undergraduate students,
who are studying band-pass signals for the first time \cite{gonzales-velasco}.\\
On the other hand, the Fourier transform, which is a widely known useful tool in different areas
such as Physics, Differential Equations, Signal Processing, etc., also has cases in which the transformation
integral needs to be computed as a principal value. One of these cases is the Fourier transform of the  function
$f(x)=\frac{1}{x}$, $x\neq 0$;

\begin{equation}\label{eq:fourier}
\mathcal{F}\left\{\frac{1}{x}\right\}(\omega)=\int\limits_{-\infty}^\infty \frac{e^{-i \omega x}}{x} dx.
\end{equation}

In the area of Signal Processing both integrals (\ref{eq:hilbert}) and (\ref{eq:fourier}) are manipulated and calculated
by clever methods which are based on convolution and inverse Fourier transform of other known functions.
Some textbooks of Complex Analysis such as \cite{soares, churchill} propose the computation of (\ref{eq:fourier}) as
an application of the theory of Residues.

In this work we are proposing a new approach for the teaching and dealing of the transformation
integrals (\ref{eq:hilbert}) and (\ref{eq:fourier}). Our treatment will be based in two already known tools of Complex analysis:
\begin{itemize}
 \item the next Proposition \ref{prop:continuidadbis} and,
 \item the Cauchy-Goursat theorem.
\end{itemize}
As far as we know, next Proposition \ref{prop:continuidadbis} was under-appreciated in the complex
analysis literature. It appears in few places like \cite{churchill} where it is posed as an exercise to be solved by using residues technique.
This Proposition helps us to solve the poles problem of the integrals (\ref{eq:hilbert}) and (\ref{eq:fourier}).
We will show it and its simple version Proposition \ref{prop:continuidad} with continuity hypothesis only, without residues theory.
On the other hand, the Cauchy-Goursat theorem synthesized by the next equation (\ref{eq:cauchy_goursat}) it is easy to understand, specially
to students who are familiar with conservative fields taught in basic courses of Physics and Calculus.\\
In order to present this new approach we organized this work as follows:\\
In Section 2 it is stated and shown that the line integral of continuous complex functions along circular arcs and around to any
point $w \in \mathbb{R}$ approaches continuously to $kf(w)$ as the radius of the arc goes to zero, where
$k$ is a constant which depends on the angle of the arc.\\
In Section 3 it is applied both the result of Section 2 and the Cauchy-Goursat theorem to study the behaviour of
principal value integral $\int_{\mathbb{R}} \frac{f(x)}{x-w} dx$ of analytical functions $f(z)$ on horizontal half planes.\\
In Section 4, results of the analysis of $\int_{\mathbb{R}} \frac{f(x)}{x-w} dx$ are applied to get a natural definition
of the Hilbert transform and also the evaluation  of the Fourier transform of the function $f(x)=\frac{1}{x}$ without
residues calculus.

\section{Approximate identities for complex continuous functions}
For real functions the Dirac's delta $\delta(t)$ is called an approximate identity in the sense that
$\int\limits_{-\infty}^\infty f(t) \delta(t-w) dt$ = $f(w)$ for any  function $f$ continuous at $w$
\cite{jackson, iorio, lathi}. Also it is known that $\delta(t)$ can be understood as the limit of a sequence of functions $\varphi_n(t)$
with $\int\limits_{-\infty}^{\infty} \varphi_n(t)dt$=1, for each $n$, in such a way that
\begin{multline}
f(w)=\int\limits_{-\infty}^\infty f(t) \delta(t-w)dt= \\
\lim\limits_{n\to \infty}\int\limits_{-\infty}^\infty f(t) \varphi_n(t-w)dt.
\end{multline}
For  complex functions we will obtain some kind of Dirac's delta
as the limit of a family of arcs  $C_r$ : $w+re^{it}$, $\theta_1\leq t\leq \theta_2$,
in the sense that
\begin{equation}\label{eq:complex_dirac}
f(w)=\lim\limits_{r\to 0} \int\limits_{\theta_1}^{\theta_2} f(C_r)\varphi(C_r-w) dC_r,
\end{equation}
where $\varphi(z)=\frac{1}{i(\theta_2-\theta_1)z}$.

\begin{center}
\begin{figure}[h]
 \includegraphics[width=6cm,scale=1]{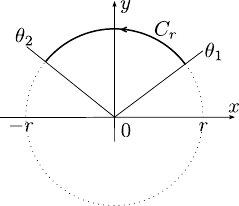}
 \caption{The  path $C_r$ : $\gamma_r(t)=re^{it}$, $\theta_1 \leq t \leq \theta_2$.}
 \label{fig:simple_path}
 \end{figure}
\end{center}

For the sake of clearness we begin to show (\ref{eq:complex_dirac}) in the simple case $w=0$.
\begin{proposition}\label{prop:continuidad}
Let $f$ be a continuous complex function in an open neighborhood around the point $z=0$.
Let $C_r$ be the path on the arc defined by  $\gamma_r(t)=re^{it}$, $t\in [\theta_1,\theta_2]$, $0 \leq \theta_1 < \theta_2\leq 2\pi$,
Figure \ref{fig:simple_path}. Then
\[\lim\limits_{r\to 0} \int\limits_{C_r} \frac{f(z)}{z}dz = i\Delta \theta f(0),\]
where $\Delta \theta=\theta_2-\theta_1$.
\end{proposition}

\textbf{Proof.- }
By the continuity of $f$ at the point $z=0$,
\[\lim\limits_{r\to 0} f(re^{it})=f(0).\]

This means that for any $\epsilon> 0$, there is some $r_0>0$ such that
$r < r_0$ implies $\vert f(re^{it})-f(0)\vert < \frac{\epsilon}{\Delta\theta}$.

On the other hand;
\begin{multline*}
\int_{C_r} \frac{f(z)}{z}dz=\int\limits_{\theta_1}^{\theta_2} \frac{f(\gamma_r(t)).\gamma_r^{\prime}(t)}{\gamma_r(t)}dt\\
=\int_{\theta_1}^{\theta_2} \frac{f(re^{it}).ire^{it}}{re^{it}} dt = i\int\limits_{\theta_1}^{\theta_2} f(re^{it}) dt
\end{multline*}

Then, writing $i\Delta\theta f(0)=i \int\limits_{\theta_1}^{\theta_2} f(0) dt$ , for  any $r<r_0$ we have
\begin{multline*}
\left \lvert \int\limits_{C_r} \frac{f(z)}{z}dz - i\Delta\theta f(0)  \right \rvert =\\
\left \lvert i\int\limits_{\theta_1}^{\theta_2} f(re^{it}) dt - i \int\limits_{\theta_1}^{\theta_2} f(0) dt \right \rvert =\\
\left \lvert \int\limits_{\theta_1}^{\theta_2} f(re^{it})-f(0) dt \right\rvert \leq \\
\int\limits_{\theta_1}^{\theta_2} \left \lvert f(re^{it})-f(0)\right \rvert dt < \\
< \left ( \frac{\epsilon}{\Delta \theta}\right ) \Delta\theta=\epsilon,
\end{multline*}
which shows that $\int\limits_{C_r} \frac{f(z)}{z}dz \rightarrow i\pi f(0)$ as $r\rightarrow 0$. $\hfill \square$

We can extend the above Proposition \ref{prop:continuidad}, which is about an arc with center $x=0$, to
an arc with arbitrary center $w\in \mathbb{R}$.
For that let $C_{wr}$ be the path over the arc defined by $\gamma_{wr}(t)=w+re^{it}$, $t\in[\theta_1,\theta_2]$ and $w$
is any real number.
Clearly, $\gamma_{wr}^\prime(t)=ire^{it}$ and
$\frac{f(\gamma_{wr}(t)).\gamma_{wr}^\prime(t)}{\gamma_{wr}(t)}=if(\gamma_{wr}(t))$. By  continuity,
$f(\gamma_{wr}(t))\to f(w)$ as $r\to 0$. Therefore, the Proposition \ref{prop:continuidad} is generalized
to the following:
\begin{propositionbis}\label{prop:continuidadbis}
Let $f$ be a continuous complex function in an open neighborhood around $z=w+i0$,  $w\in \mathbb{R}$.
Let $C_{wr}$ be the arc defined by $\gamma_{wr}(t)=w+re^{it}$, $t\in[\theta_1,\theta_2]$, then;
\[
\lim\limits_{r\to 0} \int\limits_{C_{wr}} \frac{f(z)}{z-w}dz = i\Delta\theta f(w)
\]
\end{propositionbis}
In particular for $\Delta\theta=\pi$ we have
\begin{equation}\label{continuidad}
\lim\limits_{r\to 0} \int\limits_{C_{wr}} \frac{f(z)}{z-w}dz = i \pi f(w).
\end{equation}
which means $g(z)=\frac{1}{i\pi z}$ is some kind of Dirac's delta over arcs with angle $\pi$. This result
is posed in \cite{churchill} as an application of Residues theory and without any
mention to  Dirac's delta analogy.

\section{The action of the Cauchy-Goursat theorem on horizontal half planes}
For any analytical (holomorphic) complex function $f$ with domain $D$, the
Cauchy-Goursat theorem states that
\begin{equation}\label{eq:cauchy_goursat}
\oint\limits_{C} f(z)dz=0,
\end{equation}
for any closed path $C$ inside $D$.
We will use the Cauchy-Goursat theorem together the approximate identity Propositions \ref{prop:continuidad} and \ref{prop:continuidadbis}
to prove the next Propositions \ref{prop:analytical} and \ref{prop:analyticalbis} over the upper half plane and
the Proposition \ref{prop:analytical_lower} over the lower half plane.
 These results also are shown, among others, in \cite{churchill, soares, ahlfors}, as an application of residues theory.
In our case we will use the identity approximate approach instead of residues.

\begin{center}
\begin{figure}
 \includegraphics[width=8cm,scale=1]{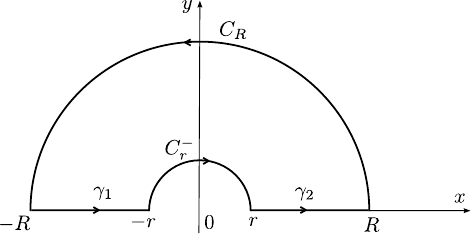}
 \caption{The closed path $\gamma=C_R*\gamma_1*C_r^-*\gamma_2$}
 \label{fig:upper_path}
 \end{figure}
\end{center}

\subsection{Cauchy-Goursat  on the upper half plane}

\begin{proposition}\label{prop:analytical}
Let $C_R$ be the semicircle $\gamma_R(t)=Re^{it}$, $t\in[0,\pi]$ and $R>0$, Figure \ref{fig:upper_path}.
 Given a complex function $f(z)$, analytical in the upper half plane $\{y\geq 0\}$, such that $f(\gamma_R(t))$ goes to zero
 as $R$ goes to infinity, denote by $f(x)$ the  restriction of $f$ to the real axis $z=x+i0$, then:
 \begin{equation}\label{eq:impropia1}
 \int\limits_{-\infty}^\infty \frac{f(x)}{x} dx= i \pi f(0)
 \end{equation}
\end{proposition}
\textbf{Proof.- }
Consider the closed path $\gamma=C_R*\gamma_1*C_r^-*\gamma_2$ of the Figure \ref{fig:upper_path}, where
$C_r^-$ is the semicircle whose parametric formula is $\gamma_r^-(t)=re^{-it}$, $t\in [\pi,2\pi]$.
Notice that the path $C_r^-$ is the orientation reversed version of the path $C_r$ from Proposition \ref{prop:continuidad}.
The parametric formulation of the paths $\gamma_1$ and $\gamma_2$ have no importance because
when $R \to \infty$ and $r\to 0$ the union of  paths $\gamma_1\cup \gamma_2$ goes to the set $\mathbb{R}-\{0\}$
which is related to the improper integral (\ref{eq:impropia1}) in the sense that
\[\int\limits_{\mathbb{R}-\{0\}} \frac{f(x)}{x}dx=\int\limits_{\mathbb{R}} \frac{f(x)}{x}dx=
\int\limits_{-\infty}^\infty \frac{f(x)}{x}dx.\]

The function $\frac{f(z)}{z}$ is analytical inside $\gamma$ then by the Cauchy-Goursat theorem:
\[\oint\limits_{\gamma} \frac{f(z)}{z}dz =0, \]
which means
\begin{equation}\label{eq:parts_closed}
\int\limits_{C_R}\frac{f(z)}{z}dz+\int\limits_{\gamma_1}\frac{f(z)}{z}dz+\int\limits_{C_r^-}\frac{f(z)}{z}dz+\int\limits_{\gamma_2}\frac{f(z)}{z}dz=0.
\end{equation}
Now, we analyze each one of the integrals of (\ref{eq:parts_closed}):\\
By the hypothesis about $f$,
\begin{equation}\label{eq:CR}
 \int\limits_{C_R}\frac{f(z)}{z}dz =
i\int\limits_0^{\pi} f(Re^{it})dt \to 0;\; \mbox{when}\; R \to \infty.
\end{equation}

On the other hand, by the relation between $C_r^-$ and $C_r$, where $C_r$ is as Figure \ref{fig:simple_path}, we have:
\[\int\limits_{C_r^-} \frac{f(z)}{z}dz = -\int\limits_{C_r} \frac{f(z)}{z}=-i\int\limits_0^{\pi} f(re^{it})dt.\]
Hence, by Proposition \ref{prop:continuidad};
\begin{equation}\label{eq:Cr}
\int\limits_{C_r^-} \frac{f(z)}{z}dz \to -i\pi f(0);\; \mbox{when}\; r\to 0.
\end{equation}
Finally, the integrals $\int\limits_{\gamma_1} \frac{f(z)}{z}dz$ and $\int\limits_{\gamma_2} \frac{f(z)}{z}dz$ go to
the integrals $\int\limits_{-\infty}^0 \frac{f(x)}{x}dx$ and  $\int\limits_0^\infty \frac{f(x)}{x}dx$, respectively,
when both $R\to \infty$ and $r\to 0$.
Hence,
\begin{equation}\label{eq:gamma12}
\int\limits_{\gamma_1} \frac{f(z)}{z}dz + \int\limits_{\gamma_2} \frac{f(z)}{z}dz \to \int\limits_{-\infty}^\infty \frac{f(x)}{x}dx
\end{equation}
Therefore, the application of (\ref{eq:CR}), (\ref{eq:Cr}), (\ref{eq:gamma12}) on the equation (\ref{eq:parts_closed})
leads us to conclude that:
\[\int\limits_{-\infty}^\infty \frac{f(x)}{x} dx= i\pi f(0).\]
$\hfill \square$\\

In the same way that the generalization of Proposition \ref{prop:continuidad} to Proposition \ref{prop:continuidadbis}
was made, we can generalize the Proposition \ref{prop:analytical} as follows;

\begin{propositionbis}\label{prop:analyticalbis}
For any $w\in \mathbb{R}$, let $C_{wR}$ be the semicircle $\gamma_{wR}(t)=w+Re^{it}$, $t\in[0,\pi]$.
 Given a complex function $f(z)$, analytical in the upper half plane $\{y\geq 0\}$, such that $f(\gamma_{wR}(t))$ goes to zero
 as $R$ goes to infinity, denote by $f(x)$ the  restriction of $f$ to the real axis $z=x+i0$, then:
 \begin{equation}\label{eq:impropia2}
 \int\limits_{-\infty}^\infty \frac{f(x)}{x-w} dx= i\pi f(w).
 \end{equation}
\end{propositionbis}

\subsection{Cauchy-Goursat on the lower half plane}

\begin{center}
\begin{figure}
 \includegraphics[width=8cm,scale=1]{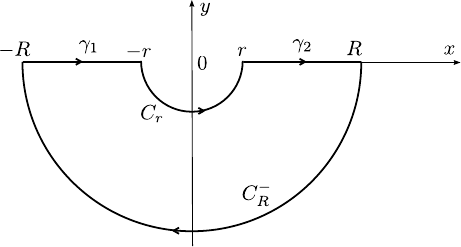}
 \caption{The closed path $\gamma=C_R^-*\gamma_1*C_r*\gamma_2$}
 \label{fig:lower_path}
 \end{figure}
\end{center}

It is important to notice that the above Propositions \ref{prop:analytical} and \ref{prop:analyticalbis} are not valid for
any analytical function. For instance the function $f(z)=z$ is analytical,
but it is false that $\int\limits_{-\infty}^\infty \frac{x}{x-w} dx= i\pi f(w) =i\pi w$.
The crucial hypothesis  in both the Propositions \ref{prop:analytical} and \ref{prop:analyticalbis} is that
$f(C_R)$, where $C_R$ is as Figure \ref{fig:upper_path}, must go to zero as $R\to \infty$.
And the function $f(z)=z$ does not satisfy this condition. A function which fully satisfies the hypotheses of both the
Propositions \ref{prop:analytical} and \ref{prop:analyticalbis} is $f(z)=e^{iz}$ and we will use it to give an important example
of Hilbert transform.\\
On the other hand if we consider the analytical function $f(z)=e^{-iz}$, its restriction over the same path $C_R$
of the Figure \ref{fig:upper_path} does not go to zero as $R\to \infty$. Instead, if we consider the path
$C_R^-$ : $\gamma_R^-{t}=Re^{-it}$, $t\in [0,\pi]$, Figure \ref{fig:lower_path},
which has clockwise orientation and it is inside the lower half plane, we have:

\begin{multline}\label{eq:lower_path}
\exp\left(-i\gamma_R^{-}(t)\right) = \exp(-iRe^{-it})=\\
\exp(-iR(\cos t-i\sin t)) = \\ e^{-iR \cos t}e^{-R \sin t},
\end{multline}
which shows that $\exp\left(-i\gamma_R^-(t)\right)$ goes to zero as $R\to \infty$.

This gives us a clue to formulate the action of the Cauchy-Goursat theorem and Proposition \ref{prop:continuidadbis}
over the lower half plane in an analogous way
the Proposition \ref{prop:analyticalbis}.

\begin{proposition}\label{prop:analytical_lower}
For any $w\in \mathbb{R}$, let $C_{wR}^-$ be the semicircle $\gamma_{wR}^-(t)=w+Re^{-it}$, $t\in[0,\pi]$.
 Given a complex function $f(z)$, analytical in the lower half plane $\{y\leq 0\}$, such that $f(\gamma_{wR}^-(t))$ goes to zero
 as $R$ goes to infinity, denote by $f(x)$ the  restriction of $f$ to the real axis $z=x+i0$, then:
 \begin{equation}\label{eq:impropia3}
 \int\limits_{-\infty}^\infty \frac{f(x)}{x-w} dx= -i\pi f(w).
 \end{equation}
\end{proposition}

\textbf{Proof.- } For simplicity let us consider $w=0$.
Consider the clockwise oriented closed path  $\gamma=C_R^-*\gamma_1*C_r*\gamma_2$ of the Figure \ref{fig:lower_path},
where $C_R$ : $\gamma_R^-(t)=Re^{-it}$, $t\in [0,\pi]$; the path $C_r$ : $\gamma_r(t)=re^{it}$, $t\in [\pi,2\pi]$ and
$\gamma_1$, $\gamma_2$ are exactly as Proposition \ref{prop:analytical}.
The function $\frac{f(z)}{z}$ is analytical inside the closed path
$\gamma$, calling Cauchy-Goursat we have:
\[\oint\limits_\gamma \frac{f(z)}{z}dz=0\]
By hypothesis,
\begin{equation}
 \int\limits_{C_R^{-}} \frac{f(z)}{z}dz \to 0;\;\mbox{when};R\to \infty.
\end{equation}
On the other
hand, by Proposition \ref{prop:continuidad}, the integral
 $\int\limits_{C_r} \frac{f(z)}{z}dz$ tends to $i\pi f(0)$
as $r$ tends to zero. Therefore
\begin{equation}\label{eq:lower0}
 \int\limits_{-\infty}^\infty \frac{f(x)}{x}=-i\pi f(0)
\end{equation}

and for an arbitrary real $w$
\begin{equation}\label{eq:lowerw}
 \int\limits_{-\infty}^\infty \frac{f(x)}{x-w}=-i\pi f(w)
\end{equation}

$\hfill \square$

\section{Applications}
\subsection{On the Hilbert transform (upper half plane)}
The signals encountered in wireless communications are typically real pass-band signals \cite{lapidoh}.
The Hilbert transform is a fundamental tool to  provide the mathematical basis for transforming
real-valued band-pass signals and systems into their low-pass equivalent representations without loss of information
\cite{haykin,lathi}.
\begin{definition}
 The Hilbert transform of a real and integrable function $u(x)$ is defined as the principal value of the following integral;
 \begin{equation}
  \mathcal{H}\{u\}(w)=\frac{1}{\pi}\int\limits_{-\infty}^\infty \frac{u(x)}{u-w} dx
 \end{equation}

\end{definition}
At first, this definition of  Hilbert transform seems artificial, without any purpose.
Also, the actual calculation of this integral seems to demand some of the most formidable techniques of calculus.
By using the Proposition \ref{prop:analyticalbis} we will see, in a natural way, that the Hilbert transform of
a real function $u(x)$ is exactly  its \textbf{harmonic conjugate}. For that, it is necessary to recall that
given an analytical function $f(z)=u(z)+iv(z)$, both of the functions $u(z)$ and $v(z)$ are called harmonic conjugates because
their partial derivatives need to satisfy both the harmonic condition of Laplace (zero Laplacian) and the
Cauchy-Riemann conditions. Detailed explanation of these ideas it can be seen in  \cite{rudin2,churchill,ahlfors,mathews}, among others.
One previous work concerned with a more natural definition of the Hilbert transform was \cite{gonzales-velasco}.

Let $f(z)=u(z)+iv(z)$ be the function satisfying the Proposition \ref{prop:analyticalbis}, then the equation
(\ref{eq:impropia2}) becomes
 \begin{multline}\label{eq:impropia4}
 \int\limits_{-\infty}^\infty \frac{u(x)+iv(x)}{x-w} dx= i\pi (u(w)+iv(w))=\\
 -\pi v(w)+ i\pi u(w).
 \end{multline}
Equating the real and imaginary parts of (\ref{eq:impropia4}) we obtain:
\begin{equation}\label{eq:hilbert1}
v(w)=\frac{1}{\pi}\int\limits_{-\infty}^\infty \frac{u(x)}{w-x}=\mathcal{H}\{u\}(w)
\end{equation}
and
\begin{equation}\label{eq:hilbert2}
u(w)=-\frac{1}{\pi}\int\limits_{-\infty}^\infty \frac{v(x)}{w-x}=-\mathcal{H}\{v\}(w).
\end{equation}
\begin{example}
 The Hilbert transform of $\cos(x)$ is $\sin(x)$ and reciprocally the Hilbert transform of $\sin(x)$ is $-\cos(x)$.
\end{example}
In effect, the complex function $f(z)=e^{iz}$ satisfy both the  hypothesis of Proposition \ref{prop:analyticalbis}.
It is clearly analytical in the whole plane and its restriction on the path $C_R$ is
 $f(\gamma_{wR}(t))=\exp(i(z+Re^{it}))$ = $ e^{i(w+R\cos t)}e^{-R\sin t}$.
 Since $\sin(t)\geq 0$ for all $t\in [0,\pi]$, then $f(\gamma_{wR}(t))$ goes to zero as $R$ goes to infinity.\\
Now,
\[f(z)=e^{iz}=e^{i(x+iy)}=e^{-y}(\cos(x)+i\sin(x)),\]
which means $u(x)=u(x+i0)=\cos(x)$ and $v(x)=v(x+i0)=\sin(x)$. Applying the formulas (\ref{eq:hilbert1}) and
(\ref{eq:hilbert2}) it can  be confirmed what was already  known by the community of signal processing arena:
$\mathcal{H}\{\cos\}(x)=\sin(x)$ and $\mathcal{H}\{\sin\}(x)=-\cos(x)$.

\subsection{On the Fourier transform (lower half plane)}
It is widely known that the Fourier transform of a real function $f(x)$ is
\[\mathcal{F}\{f\}(\omega)=\int\limits_{-\infty}^\infty f(x) e^{-i\omega x} dx.\]
In particular if $f(x)=\frac{1}{x}$ and the frequency is $\omega=1$  we have
\[\mathcal{F}\left\{\frac{1}{x}\right\}(\omega)\big\vert_{\omega=1}=\int\limits_{-\infty}^\infty \frac{e^{-i x}}{x} dx.\]
As it was said before, the function $f(z)=e^{-iz}$ satisfy the hypotheses of Proposition \ref{prop:analytical_lower}, therefore
\begin{equation}\label{eq:fourier1}
\mathcal{F}\left\{\frac{1}{x}\right\}(\omega)\big\vert_{\omega=1}=\int\limits_{-\infty}^\infty \frac{e^{-i x}}{x} dx=-i\pi.
\end{equation}
 With this, we can find the Fourier transform  for any
positive frequency $\omega>0$ by doing a simple change of integration variable $x=\omega u$, then;
\[\int\limits_{-\infty}^\infty \frac{e^{-ix}}{x}dx= \int\limits_{-\infty}^\infty \frac{e^{-i\omega u}}{\omega u}d(\omega u)
\int\limits_{-\infty}^\infty \frac{e^{-i\omega u}}{u}du.\]
In this way it has been found  another way to compute the Fourier transform of the function $f(x)=\frac{1}{x}$:

\[\mathcal{F}\left\{\frac{1}{x}\right\}(\omega)=\int\limits_{-\infty}^\infty \frac{e^{-i\omega x}}{x}=\begin{cases}
                             -i\pi;\;\mbox{if}\; \omega>0\\
                             i\pi;\;\mbox{if}\; \omega<0
                            \end{cases}
\]

Other interesting consequence of equation (\ref{eq:fourier1}) is
\[\int\limits_{-\infty}^\infty \frac{\sin(x)}{x} dx =\pi.\]

\section{Conclusions}
In this work we had shown that there is a way to compute some principal value integrals without residues theory.
The tool  we use instead of residues calculus is a result that has some resemblance to the Dirac delta which was called
approximate identity. This result is synthesized as the Proposition \ref{prop:analyticalbis}  only requires $f(z)$
be continuous around a real point and it was under-appreciated in the complex analysis literature.\\
With our approach, it is clearly understood that the Hilbert transform is not an artificial formula but that
the pair $u(x)$ and $v(x)=\mathcal{H}\{u\}(x)$ are the restriction on the real axis of an analytical function
$f(z)=u(z)+iv(z)$.


\begin{thebibliography}{99}


\bibitem{haykin}
S.~Haykin, \emph{Digital Communication Systems}.\hskip 1em plus 0.5em minus
  0.4em\relax New York: Wiley and Sons, 2014.

\bibitem{lathi}
B.~Lathi and Z.~Ding, \emph{Modern Digital and Analog Communication Systems},
  5th~ed.\hskip 1em plus 0.5em minus 0.4em\relax New York, NY USA: Oxford
  University press, 2019.

\bibitem{zayed}
A.~Zayed, ``Hilbert transform associated with the fractional fourier
  transform,'' \emph{IEEE Signal Processing Letters}, vol.~5, pp. 206--208,
  1998.

\bibitem{brackx}
H.~D.~S. Fred~Brackx, Bram De~Knock and D.~Eelbode, ``On the interplay between
  the hilbert transform and conjugate harmonic functions,'' \emph{Mathematical
  Methods in the Applied Sciences}, vol.~29, pp. 1435--1450, 2006.

\bibitem{caleb}
C.~Marshall, ``A field guide for hilbert transforms with new estimates on an
  associated maximal directional operator,'' Master's thesis, University of
  British Columbia, Vancouver, 2021.

\bibitem{gonzales-velasco}
E.~Gonzales-Velasco and E.~Sanvicente, ``The analytic representation of
  band-pass signals,'' \emph{Journal of The Franklin Institute}, vol. 310, pp.
  135--142, 1980.

\bibitem{soares}
M.~G. Soares, \emph{Cálculo em uma Variável Complexa}, quinta~ed.\hskip 1em
  plus 0.5em minus 0.4em\relax Rio de Janeiro: IMPA, 2016.

\bibitem{churchill}
J.~Brown and R.~Churchill, \emph{Complex Variables and Applications},
  9th~ed.\hskip 1em plus 0.5em minus 0.4em\relax McGraw Hill, 2013.

\bibitem{jackson}
J.~D. Jackson, \emph{Classical Electrodynamics}, 3rd~ed.\hskip 1em plus 0.5em
  minus 0.4em\relax new York: Wiley and Sons, 1999.

\bibitem{iorio}
R.~Iorio and V.~Iorio, \emph{Equações Diferenciais Parciais: Uma
  Introdução}, terceira~ed.\hskip 1em plus 0.5em minus 0.4em\relax Rio de
  Janeiro: IMPA, 2018.

\bibitem{ahlfors}
L.~Ahlfors, \emph{Complex Analysis: An Introduction to the Theory of Analytic
  Functions of One Complex Variable}, 3rd~ed.\hskip 1em plus 0.5em minus
  0.4em\relax AMS-Chelsea Publishing, 2021.

\bibitem{lapidoh}
A.~Lapidoth, \emph{A Foundation in Digital Communication}.\hskip 1em plus 0.5em
  minus 0.4em\relax Cambridge, UK: Cambridge University Press, 2009.

\bibitem{rudin2}
W.~Rudin, \emph{Real and Complex Analysis}, 3rd~ed.\hskip 1em plus 0.5em minus
  0.4em\relax New York: McGraw-Hill, 1986.

\bibitem{mathews}
J.~Mathews and R.~Howell, \emph{Complex Analysis for Mathematics and
  Engineering}, 6th~ed.\hskip 1em plus 0.5em minus 0.4em\relax Jones and
  Bartlett, 2012.

\end{thebibliography}
\end{document}